# Nonreciprocal Charge Transport in Quantum Materials: From Superconductors to Magnetic Systems

**Boning Yu**[1,*], **Yalei Huang**[1], **Alltrin Dhanarajagopal**[1], **Xiaofeng Xu**[1,*]

[1] School of Physics and Optical Engineering, Zhejiang University of Technology, Hangzhou 310023, China

*Author to whom any correspondence should be addressed.

**E-mail:**

boningyu@zjut.edu.cn (B. Y.)

xuxiaofeng@zjut.edu.cn (X. X.)

**Nonreciprocal charge transport, where current flows preferentially in one direction, has become a key topic across quantum materials. This review covers some recent progress in bulk crystals, artificial heterostructures, van der Waals systems, high-temperature superconductors, and magnetic materials. The review discusses both intrinsic origins, such as band-structure effects, and extrinsic ones, including vortex motion, device geometry, and heating artifacts. A central focus is the superconducting diode effect due to its application potential, which shows different critical currents for opposite directions. The review also examines how symmetry breaking, spin–orbit coupling, and magnetic order contribute to the observed responses. However, there remain challenges when many exciting findings lack a unified picture, and the field continues to debate the relative importance of various mechanisms. The review highlights key open questions and suggests future directions, including better experimental standards and device integration. This review aims to give a balanced view of what is known, what is uncertain, and where the field might go next.**

## 1. Introduction

Among the usual electronic components such as diodes and transistors, electric current flows freely in one direction but is prevented in the opposite direction. This direction-dependent behavior is known as nonreciprocity. In semiconductors, this property is achieved by the design of a p–n junction, where a hole-rich p-type region meets an electron-rich n-type region. The difference in carrier concentration creates a built-in electric field at the interface, which rectifies the current. Common materials like Si, Ge, or GaAs are lightly doped with atoms that have fewer or more electrons to realize such junctions.

Over the past two decades, researchers have discovered that certain materials exhibit nonreciprocal transport intrinsically, without any artificial junction. This has attracted growing interest to condensed matter physics, as it offers new ways to study symmetry breaking and may lead to novel device concepts. However, a major challenge in the field is distinguishing between intrinsic nonreciprocity (i.e. the electronic band structure) and extrinsic mechanisms (i.e. vortex edge barriers, geometric asymmetries, or thermoelectric artifacts). Many claims of intrinsic effects have later been questioned by alternative explanations, and this debate remains central to the field.

According to Onsager's reciprocity theorem [1], a linear nonreciprocal electrical response requires both time-reversal symmetry (TRS) and inversion symmetry (IS) to be broken. In transport experiments, the resistance $R$ of a noncentrosymmetric conductor under a magnetic field $B$ and current $I$ is often written as [2-4]

$$R = R_0(1 + \beta B^2 + \gamma BI)$$

where $R_0$ is the zero-field resistance, $\beta$ describes ordinary magnetoresistance, and $\gamma$ is the nonreciprocal coefficient. When either the magnetic field or the current direction is reversed, the $\gamma BI$ term changes sign, leading to an asymmetric resistance with respect to current reversal.

In recent years, the idea of nonreciprocal transport has been extended to superconductors. There, nonreciprocity appears as a difference between the critical currents for opposite directions, $I_{c+} \neq I_{c-}$, a phenomenon known as the superconducting diode effect (SDE) [5-8]. The efficiency of this diode is defined as

$$\eta = \frac{I_{c+} - |I_{c-}|}{I_{c+} + |I_{c-}|} \times 100\%$$

When $\eta = 100\%$, the device acts as an ideal rectifier: superconductivity survives for one current direction but is fully destroyed for the other [9]. In practice, however, most reported efficiencies are below 20%, and achieving high efficiency often reduces the critical current or narrows the usable temperature and field range. This trade-off has not received enough attention in the literature, and we will touch on it where relevant.

Like the linear case, the SDE also requires both IS and TRS to be broken [4, 7, 10, 11]. IS can be broken by noncentrosymmetric crystal structures, heterostructure interfaces, artificial patterning, or ferroelectricity. TRS is usually achieved through magnetic order or an external magnetic field. Yet the symmetry conditions alone are not sufficient: the orientation of the magnetic field relative to the crystal axes and the current direction matters greatly. Moreover, as we discuss later, one must be cautious about interpreting SDE signals, when Joule-heating-induced thermoelectric effects can produce signals that look like nonreciprocal critical currents [12]. Recent studies on FeSe have highlighted this issue [13, 14].

In this review, we summarize some recent progress on nonreciprocal charge transport across a broad range of quantum materials. We organize the discussion by material platform, which allows us to examine how multiple microscopic mechanisms, that sometimes coexist or compete, can operate within the same family of materials. We begin with bulk crystals with broken symmetry, including chiral and kagome superconductors. Next come artificial heterostructures, such as metallic superlattices, oxide interfaces, and superconductor–ferromagnet hybrids. We then focus on layered and low-dimensional van der Waals systems, from heterojunctions to few-layer devices. The high-$T_c$ superconductors of iron-based, cuprate, and nickelate follow, with special attention to the debate between thermoelectric artifacts and genuine effects. We also survey non-superconducting topological and magnetic materials, which provide useful comparative perspectives. Finally, we close with a summary of open questions and outlooks.

## 2. Symmetry-broken bulk crystals

Single crystals are the foundation of many research topics in condensed matter, some of which with nonsymmetric structure provides a straightforward platform for studying charge motions. This intrinsic symmetry breaking property without the influence of interface or artificial designs, makes single crystals a clear target for revealing the microscopic origin of nonreciprocity.

## 2.1 Noncentrosymmetric and chiral crystals

The study of nonreciprocal charge transport in bulk single crystals has historically been anchored in materials with broken inversion symmetry [15, 16], where the Rashba or chiral band structure provides a natural platform for directional asymmetry [17]. A landmark early work by Ideue et al. [18] reported a giant bulk rectification coefficient $\gamma \sim 1\,A^{-1}T^{-1}$ in the polar Rashba semiconductor BiTeBr, quantitatively reproduced by Boltzmann transport theory based on a 3D Rashba model [2]. It established that $\gamma$ was governed solely by intrinsic band parameters and was independent of the single-relaxation-time approximation. This result set a benchmark for intrinsic nonreciprocity in noncentrosymmetric systems.

Moving beyond, Liu et al. [19] investigated the chiral superconductor $NbGe_2$ and observed a giant rectification effect during the superconducting transition, with $\gamma$ reaching as high as $6 \times 10^5\,A^{-1}T^{-1}$ (normalized $\gamma' \sim 4 \times 10^{-7} m^2A^{-1}T^{-1}$). This effect was attributed to vortex dynamics under asymmetric pinning rather than band-structure asymmetry [20], highlighting that even in chiral crystals, the microscopic origin of nonreciprocity can be extrinsic. Similarly, Park et al. [21] reported pronounced second-harmonic resistance $R_{2\omega}$ signals near the superconducting transition in chiral $Mo_3Al_2C$, with a notable persistence of finite $R_{2\omega}$ under perpendicular fields, suggesting a toroidal contribution beyond the conventional electrical magnetochiral anisotropy (eMChA). The competing enhancement and suppression mechanisms they observed underscore the complexity of disentangling intrinsic and extrinsic contributions in chiral superconductors.

A distinct approach was taken by Nakamura et al. [22] in the chiral magnet $Co_8Zn_9Mn_3$, which exhibits a helical spin structure above room temperature. They successfully identified two separate nonreciprocal contributions: one dominant near $T_c$ and in the forced ferromagnetic state, arising from asymmetric scattering due to spin fluctuations [23], and another most pronounced in the conical phase and persisting to low temperatures, attributed to intrinsic band asymmetry independent of scattering [24]. Quantitative comparison with Boltzmann theory supported this assignment, while a linear-in-B term from spin–orbit coupling was found negligibly small. This separation of scattering-dominated and band-geometric contributions provides a valuable methodological template for other systems.

Returning to superconductors, Jiang et al. [25] reported a sign-controllable zero-field SDE in the van der Waals trigonal superconductor $PtBi_2$, achieved by trapped magnetic flux that breaks TRS. Notably, the effect persisted for both in-plane field orientations (perpendicular and parallel to current) and exhibited hysteresis with respect to the poling field, while no such effect appeared under out-of-plane fields. However, the authors acknowledged that the inversion-symmetry relations to crystal polarity or shape asymmetry remain to be further identified, leaving room for alternative interpretations [26]. Taken together, these bulk-crystal studies reveal that even within the "intrinsic" category, the operative mechanisms span band geometry, vortex dynamics, spin fluctuations, and flux trapping—a diversity that calls for case-by-case scrutiny rather than a unified narrative.

## 2.2 Kagome superconductors

Kagome superconductors have emerged as an especially fertile platform for studying nonreciprocal transport physics, owing to their unique lattice structures, topological bands, and rich symmetry-breaking orders. Fig.1a presents the centrosymmetric layer of $AV_3Sb_5$ (A = K, Rb, Cs) structure, where an in-domain chirality can be spontaneously developed as loop currents, as illustrated in Fig.1b. In $CsV_3Sb_5$ that attracted most interest among kagome superconductors so far, Guo et al. [27] demonstrated chiral transport in low-strain microdevices with the current along c-axis, where nonreciprocal signals were induced only at a small-angle tilted in-plane magnetic field, as shown in Fig.1c. However, as depicted in Fig.1d, a comparative study [28] by the same group found that $KV_3Sb_5$ in the same conditions, despite having similar lattice and band structures, failed to show obvious nonreciprocal features. Furthermore, Fig.1e provided a more common scenario in $CsV_3Sb_5$ flakes where the driving current was applied along the ab-plane, Wu et al. [29] observed multi-peaked second-harmonic resistance in the superconducting transition with an external out-of-plane magnetic field, indicating a complex symmetry-broken nature. To notice, the direct nonreciprocal report on $KV_3Sb_5$ with in-plane current is comparatively limited. This contrast is striking: it suggests that nonreciprocal transport in kagome superconductors is unlikely to be determined solely by the ideal kagome band structure; instead, it appears sensitively dependent on electronic correlations, ordered states, and disorder. This conclusion calls for more systematic, material-specific theoretical treatments that go beyond generic band-structure arguments.

Theoretically, Tazai et al. [30] proposed that in the chiral loop-current phase of $CsV_3Sb_5$, the momentum dependence of the quantum metric may undergo resonant enhancement, leading to significant eMChA. Experimentally, Le et al. [31] displayed a zero-field SDE in $CsV_3Sb_5$, where they also observed double-slit superconducting interference patterns, suggesting broken TRS inside the superconducting domain structure. Uniquely in $RbV_3Sb_5$, Wang et al. [32] observed unconventional hysteresis and reentrant superconductivity at 400mK, along with SDE in irradiated samples as demonstrated in Fig.1f, further evidencing broken TRS.

Beyond transport measurements, which provide macroscopic evidence, other techniques offer microscopic clues. Scanning tunneling microscopy (STM) has revealed chiral $2 \times 2$ charge order with magnetic-field-tunable chirality in $KV_3Sb_5$ [33] and $CsV_3Sb_5$ [34], and more recently, a chiral $2 \times 2$ pair-density wave (PDW) whose chirality can be switched by field training, providing direct spectroscopic evidence for TRS breaking in the superconducting state [35]. This broken TRS is further substantiated by muon spin relaxation ($\mu SR$) measurements, which detect enhanced internal field width below the charge ordering temperature, indicative of spontaneous orbital currents [36]; in Ta-doped $CsV_3Sb_5$ where charge order is suppressed, similar $\mu SR$ experiments reveal spontaneous internal magnetism emerging exclusively below $T_c$, evidencing TRS breaking intrinsic to the superconducting state itself [37]. The presence of a PDW with finite-momentum Cooper pairing naturally leads to asymmetric vortex dynamics and a spatially modulated superconducting order parameter [35]. Furthermore, spatially resolved photoemission spectroscopy (SPEM) reveals intrinsic nanoscale inhomogeneities and percolative CDW domains, which may act as local symmetry-breaking centers that enhance directional transport anisotropy [38]. Collectively, these observations establish a strong microscopic basis for field-trainable SDE and other nonreciprocal phenomena in kagome superconductors, although the relative importance of PDW, CDW, and vortex effects in different members of the family remains an open question.

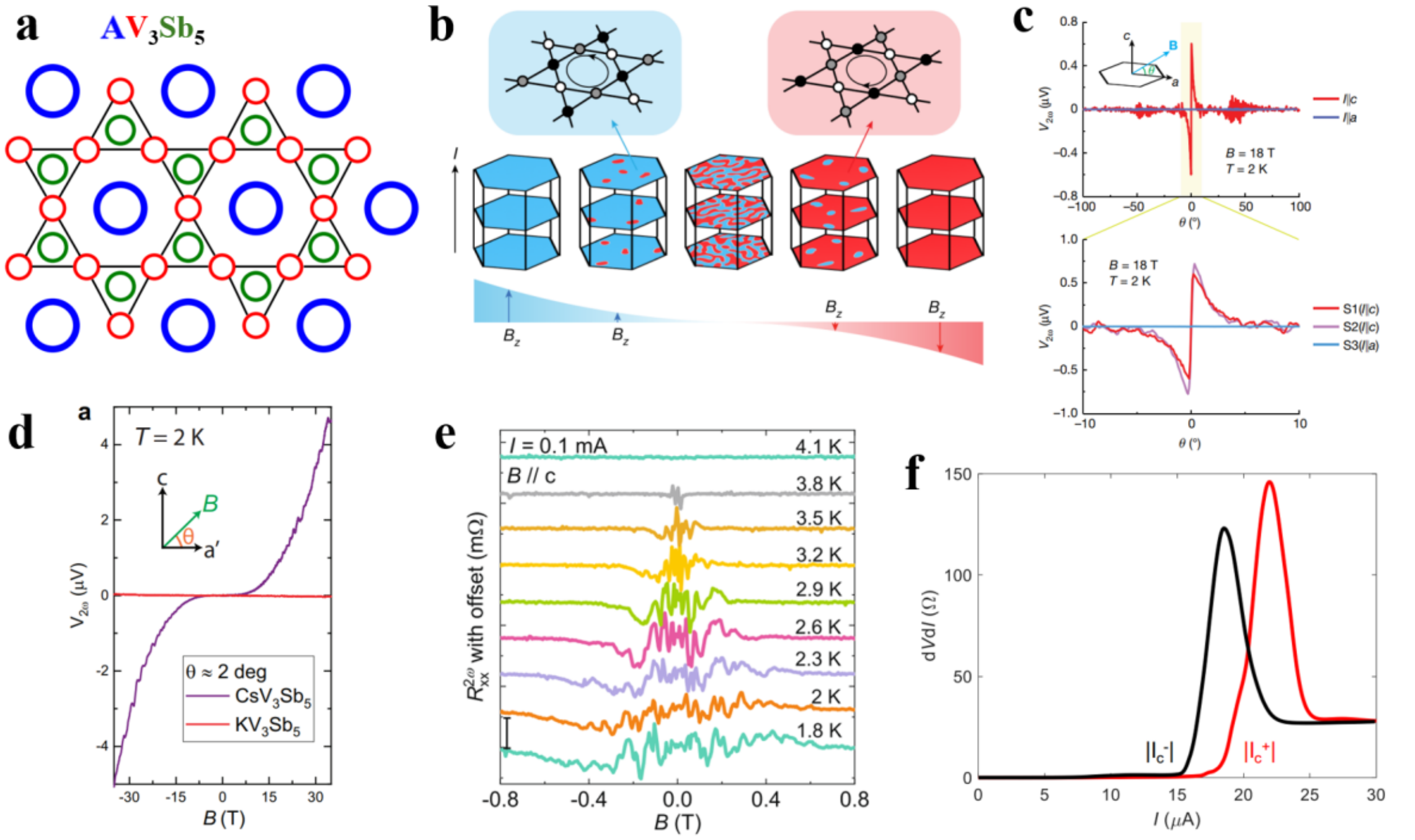

**Fig.1** Nonreciprocal transport features on kagome superconductors. a) Structure of kagome superconductors $AV_3Sb_5$ (A = K, Rb, Cs). b) Spontaneous symmetry breaking of a chiral loop current under a magnetic field at low temperature. c) Second-harmonic voltage $V_{2\omega}$ versus the angle between current and the magnetic field in $CsV_3Sb_5$ devices. d) Different second-harmonic voltage $V_{2\omega}$ responses versus a small-angled magnetic field for $CsV_3Sb_5$ and $KV_3Sb_5$, with the current along c-axis. e) The second-harmonic resistance $R_{xx}^{2\omega}$ versus an out-of-plane magnetic field for $CsV_3Sb_5$, with the current along ab plane. f) The SDE in $RbV_3Sb_5$ under irradiation. Panel (b-c) are reproduced from [27], CC BY 4.0. Panel (d) is reproduced from [28], CC BY 4.0. Panel (e) is reproduced from [29], CC BY 4.0. Panel (f) is reproduced with permission from [32], Copyright (2025), The Authors, some right reserved; exclusive licensee Springer Nature, CC BY-NC-ND 4.0.

## 3. Artificial heterostructures

Stacked or combined with different compounds in atomic scale, artificial heterojunctions offer more flexible modulations than single crystals in engineering of nonreciprocal transport.

## 3.1 Metallic superlattices

Moving from bulk crystals to artificially layered systems offers additional flexibility in tailoring symmetry breaking and spin–orbit coupling. The first experimental observation of SDE in a superlattice was reported by Ando et al. [39] in a $[Nb(1nm)/V(1nm)/Ta(1nm)]_{40}$ stack, where the effect was attributed to magnetochiral anisotropy originating from Rashba spin–orbit interaction [40], with a maximum eMChA coefficient $\gamma \sim 550\ A^{-1}T^{-1}$ under in-plane magnetic field. However, a critical re-examination by Li et al. [41] using time-dependent Ginzburg–Landau (TDGL) simulations suggested that the observed SDE in this same system may have a significant contribution from vortex dynamics. Specifically, differences in coherence lengths among elements led to material-dependent preferred vortex motion, which enhanced or reduced the SDE depending on the Lorentz force direction. This challenges the original interpretation based purely on Rashba spin–orbit coupling and highlights the difficulty of unambiguously assigning a single microscopic mechanism in artificially layered systems—a recurring theme in the field.

One forefront effort to realize zero-field SDE involves replacing the external magnetic field with an exchange field from a ferromagnetic layer. Narita et al. [42] inserted a magnetic Co layer into the $[Nb/V/Ta]_n$ multilayers and achieved a zero-field SDE with polarity reversal upon magnetization switching, demonstrating a promising route toward practical, field-free devices.

## 3.2 Oxide heterointerfaces

Oxide heterojunctions hosting two-dimensional electron gases (2DEGs) provide another versatile platform, where strong Rashba spin–orbit coupling and high carrier mobility can be tuned externally. Fig.2a shows a light-triggered nonreciprocal signal at the interface of $CaZrO_3/KTaO_3$ [43], where ultraviolet illumination boosts $\gamma$ by a factor of 1000. Fig.2b further reveals that the Rashba effective field $B_{so}$ reaches 14.7T under light of 330nm, attributing this enhancement to photoinduced electron excitations across the interband. Subsequently, Wang et al. [44] realized an SDE at the $LaAlO_3/KTaO_3$ interface under a small perpendicular field, with efficiency reaching 40% and rectification output of 10mV. Notably, a local conductive atomic force microscopy lithography (cAFM) was applied to remove the roughness on the device edge, following with the vanishing of SDE. This geometry excluded finite-momentum Cooper pairing as the origin, instead supporting a vortex edge asymmetry model [26]. In a different oxide system, a 2DEG at the Si–$SiO_2$–ZnO heterostructure was found to exhibit both nonreciprocal charge transport (NCT) and nonlinear Hall effect (NLHE) [45]. Consistent with scaling relations, NCT was attributed to Rashba spin–orbit coupling [46] and NLHE to extrinsic skew scattering [47], while a strong dependence on scattering time ruled out Berry curvature dipole and side-jump as dominant contributors. These oxide studies illustrate that even within the same broad materials class, the dominant mechanisms can differ markedly depending on interface chemistry, carrier density, and scattering regime.

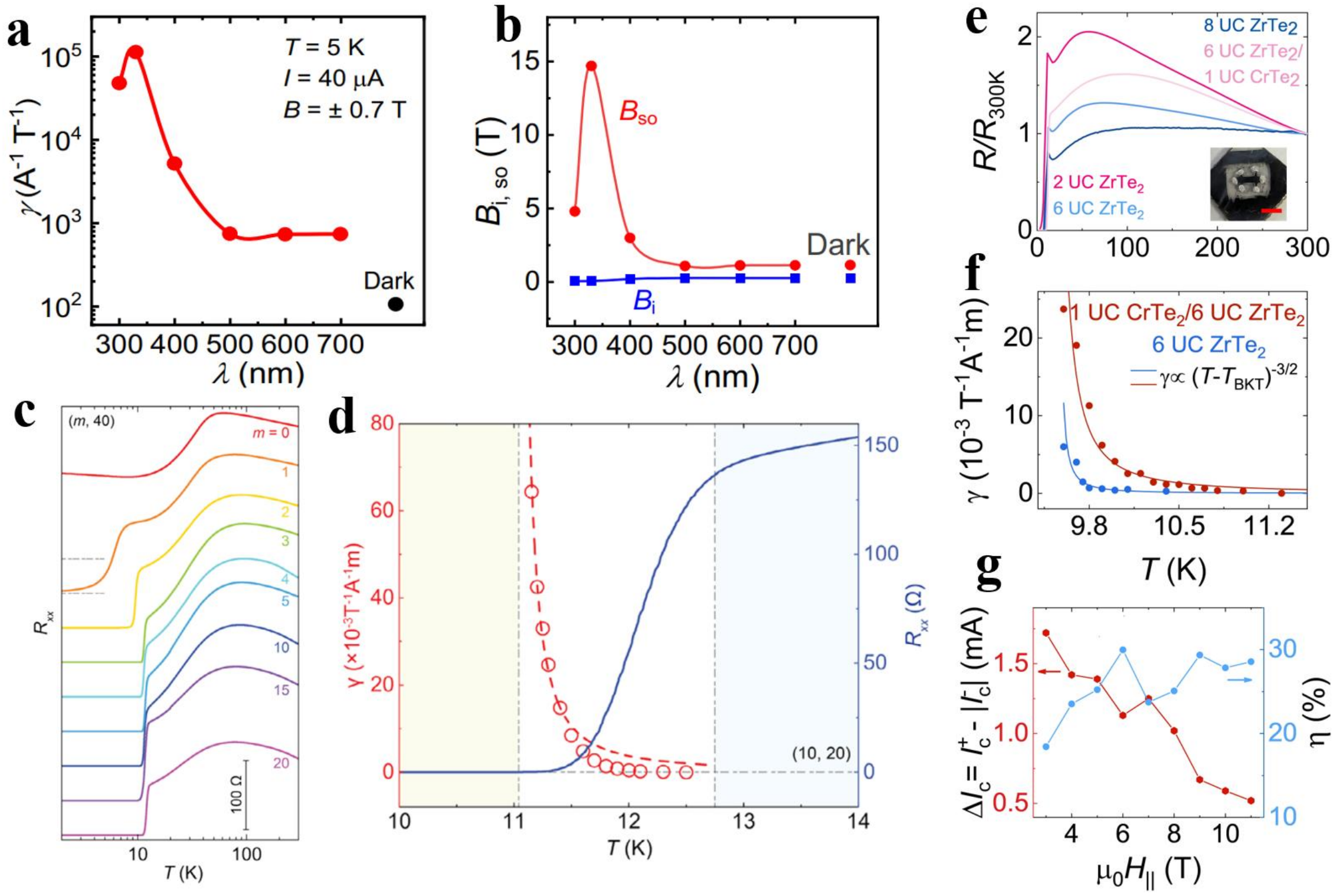

**Fig.2** Transport signals in heterojunctions. a) Nonreciprocal coefficient $\gamma$ versus the wavelengths of irradiation lights in $CaZrO_3/KTaO_3$ films. b) Rashba SOC effective field $B_{so}$ and inelastic scattering field $B_i$ versus the irradiation lights in $CaZrO_3/KTaO_3$ films. c) RT curves of $CrTe_2$/FeTe films with m trilayer of 1T-$CrTe_2$ stacking on 40 unit-cell of FeTe. d) Magneto-chiral anisotropy coefficient $\gamma$ and longitudinal resistance $R_{xx}$ versus temperature of $CrTe_2$/FeTe films with 10 trilayers of 1T-$CrTe_2$ stacking on 20 unit-cell of FeTe with BKT fitting. e) RT curves of $ZrTe_2$/FeTe and $CrTe_2/ZrTe_2$/FeTe films with different thicknesses. f) Magneto-chiral coefficients $\gamma$ with BKT fitting versus temperature at $I = 600\mu A$ for 6 UC $ZrTe_2$ films, capped and uncapped by 1UC $CrTe_2$.

g) SDE characteristics of 1UC $CrTe_2$ capped device in Fig.2g. Panels (a-b) are reproduced from [43], CC BY 4.0. Panels (c-d) are reproduced from [48], CC BY 4.0. Panels (e-g) are reproduced from [49], CC BY 4.0.

## 3.3 SC/FM hybrid structures

Hybrid structures combining superconductors with ferromagnets offer perhaps the most direct route to breaking both inversion and time-reversal symmetries simultaneously. Hou et al. [26] systematically investigated SDE in conventional superconducting V and Nb films, their bilayers with ferromagnet Eu, and trilayers $Nb/Al_2O_3/EuS$. In each structure, they verified that the underlying mechanism was asymmetric Meissner screening combined with vortex surface barriers [50, 51], rather than finite-momentum Cooper pairing. Crucially, they demonstrated that rectification arises from the superposition of applied current with Meissner screening currents generated either by external out-of-plane fields or by fringing fields from the ferromagnetic layer, and that this effect does not require spin–orbit coupling or direct exchange interaction at the SC/FM interface. This work serves as an important reminder that seemingly "exotic" SDE signals can arise from quite conventional electrodynamic effects.

In a Pt/Co/Nb heterojunction, Zhu et al. [52] demonstrated that a triangular bump on the device geometry, combined with a perpendicularly magnetized Co layer, enabled a zero-field SDE with an efficiency of 54%, resulting from the interplay of asymmetric vortex entry, spatially selective pinning caused by stray-field gradients, and unbalanced Lorentz forces.

Interface-induced superconductivity and enhanced nonreciprocal signals ($\gamma \sim 6.43 \times 10^{-2} A^{-1} T^{-1} m$) were reported in 1T-$CrTe_2$/FeTe heterostructures [48], where the response emerges predominantly within the superconducting fluctuation regime above $T_{BKT}$ and diverges as temperature approaches the BKT transition. Fig.2c shows the RT curves of multiple $CrTe_2$/FeTe structures where superconducting transition can be clearly identified, stacking with m-UC of $CrTe_2$ on n-UC of FeTe, noted as (m, n). Fig.2d shows the BKT fitting of the magentochiral anisotropy coefficient in range of superconducting transitions. Critically, the observation of such large nonreciprocal transport in a nontopological system rules out the necessity of topological surface states, pointing instead to the superconducting FeTe layer near the interface as the origin. Recently, Islam et al. [49] reported emergent 2D superconductivity and pronounced nonreciprocal transport in $ZrTe_2$/FeTe heterostructures. Fig.2e shows the resistance-temperature curves of different films, with clear superconducting transitions. In the 6 UC $ZrTe_2$ device, $\gamma$ reached $6\ A^{-1}T^{-1}m$, while further capped with 1UC $CrTe_2$, $\gamma$ even tripled and the SDE efficiency reached 29%, as illustrated in Fig.2f and 2g, respectively. The enhancement might be due to the TRS breaking by the additional ferromagnetic $CrTe_2$ layer, while the superconducting state was confirmed via BKT analysis and large upper critical fields, the origin of superconductivity and the mechanism behind nonreciprocal enhancement remain to be firmly established, and certain anomalous features would benefit from further investigation.

In a different hybrid architecture, Yu et al. [53] reported a gate-tunable SDE in planar Josephson junctions fabricated on InAs heterostructures with periodic holes etched into the Al contacts. The diode efficiency $\eta$ exhibited a nonmonotonic dependence on top-gate voltage, peaking near the depletion threshold of the holes, while junction gate tuning yielded a monotonic decrease. Numerical simulations attributed the enhancement to depletion-induced changes in transparency differences between transverse channels, rather than to chemical potential or Rashba modulation within the weak link [53]. Prior work in the same material platform further revealed that the diode efficiency can even change sign depending on contact width and field orientation [54]. The diversity of mechanisms proposed across these hybrid structures ranging from Meissner screening to stray-field pinning to interface-induced superconductivity, suggests that a unified picture remains elusive, and that careful control experiments are essential to distinguish among competing interpretations.

# 4. Layered and low-dimensional superconductors

Reducing dimensions usually comes along with the change of symmetry and the emergence of new physics phenomena, therefore layered and low-dimensional superconductors become the frontline of nonreciprocal investigations.

## 4.1 vdW heterojunctions

Van der Waals heterojunctions offer atomically sharp interfaces and the ability to combine disparate materials with different symmetries and electronic properties. The first experimental realization of a zero-field Josephson diode effect was achieved by Wu et al.

[55] in $NbSe_2/Nb_3Br_8/NbSe_2$ heterojunctions at 20mK, enabling half-wave rectification with a ratio $\sim 10^4$. The positive and negative critical currents $I_{c+}$ and $I_{c-}$ at zero field, have a difference $\Delta I_c = |I_{c-}| - I_{c+}$ that remains symmetric with magnetic field and does not reverse. In compare, the $NbSe_2/NbSe_2$ junction without the $Nb_3Br_8$ layer shows a mirrored relation of $I_{c+}$ and $I_{c-}$, and $\Delta I_c$ is antisymmetric with the field, revealing the zero-field SDE is not driven by the magnetic field [56]. In the absence of Joule heating, ferromagnetism, magnetochiral anisotropy, or electrode asymmetry, the asymmetric Josephson tunneling was attributed to an out-of-plane electric polarization arising from the obstructed atomic insulator nature of $Nb_3Br_8$ combined with structurally inequivalent top and bottom interfaces [57]. This work stands as a landmark demonstration that purely structural asymmetry at the atomic scale can yield a large zero-field diode effect.

In vertical $NbSe_2/T_d$-$WTe_2/NbSe_2$ Josephson junctions, Kim et al. [58] observed substantial supercurrent non-reciprocity under in-plane magnetic fields. Based on magneto-linearity, matching of magneto-chirality with crystal axes, and a linear efficiency-thickness relation below 23 nm, the effect was attributed to intrinsic inversion-symmetry breaking of the $T_d$-$WTe_2$ barrier. Notably, extrinsic vortex pinning, kinetic inductance, or interface asymmetries were not found in $MoTe_2$ control junctions, strengthening the case for an intrinsic mechanism.

By integrating the antiferromagnetic Mott insulator $\alpha$-$RuCl_3$ with the superconductor $NbSe_2$, He et al. [59] successfully induced superconductivity in $\alpha$-$RuCl_3$ via proximity effect, with a 0.2meV superconducting gap confirmed and a pronounced SDE with $\eta$ up to $15.4\%$ under a weak out-of-plane magnetic field—interpreted as Ising-type spin–orbit coupling rather than in-plane Rashba-type [60]. Improved from common SDEs modulated by single-direction fields, a recent dual-mode SDE was realized in a $NbS_2/NbSe_2$ heterojunction by Guan et al. [61], allowing independent induction of SDE by either out-of-plane (Fig.3a) or in-plane (Fig.3b) magnetic fields. The efficiency induced by an out-of-plane field follows a Gaussian-like peaks with maximum value at a small field $\pm$0.4mT, while the in-plane field induced SDE shows a sinusoidal pattern in the range of $\pm$300mT [62]. Both $NbS_2$ and $NbSe_2$ possess strong Ising SOC, and the reduced symmetry at the interface creates conditions for SDE under both field orientations. In comparison, the dual-mode SDE was barely detectable in a $NbSe_2/NbSe_2$ homostructure [61] despite symmetry breaking at the interface [63], further associating the dual-mode effect with the cooperative interplay of out-of-plane Ising SOC and in-plane Rashba SOC [64].

By first-principles investigations of a van der Waals ferroelectric Bi/SnTe heterojunction, Wang et al. [65] predicted an enhancement of nonlinear Hall effect and kinetic magnetoelectric effect induced by multistate polarization. These enhancements originated from ferroelectric-driven band splitting, accompanied by redistribution of Berry curvature and orbital magnetization across different polarization states [66]. Moreover, both sign and magnitude of these nonreciprocal responses were predicted to be electrically tunable by switching between eight distinct polarization configurations—offering a promising route for multifunctional nanoelectronic devices, although experimental verification remains pending.

In planar $NbSe_2/Bi_2Se_3$ Josephson junctions, Kudriashov et al. [67] reported a non-reciprocal supercurrent and a tunable JDE near a single flux quantum $\Phi_0$, with the current-phase relation exhibiting an anomalous peak-dip structure at $\Phi_0$ (equivalent to about 0.746mT), as shown in Fig.3c. The analytic results without invoking Majorana bound states in Fig.3d suggested the anomalies to originate from the interplay of a non-uniform, edge-enhancing current and non-sinusoidal local CPR, which was a useful counterpoint to proposals that attribute similar anomalies to topological superconductivity [68].

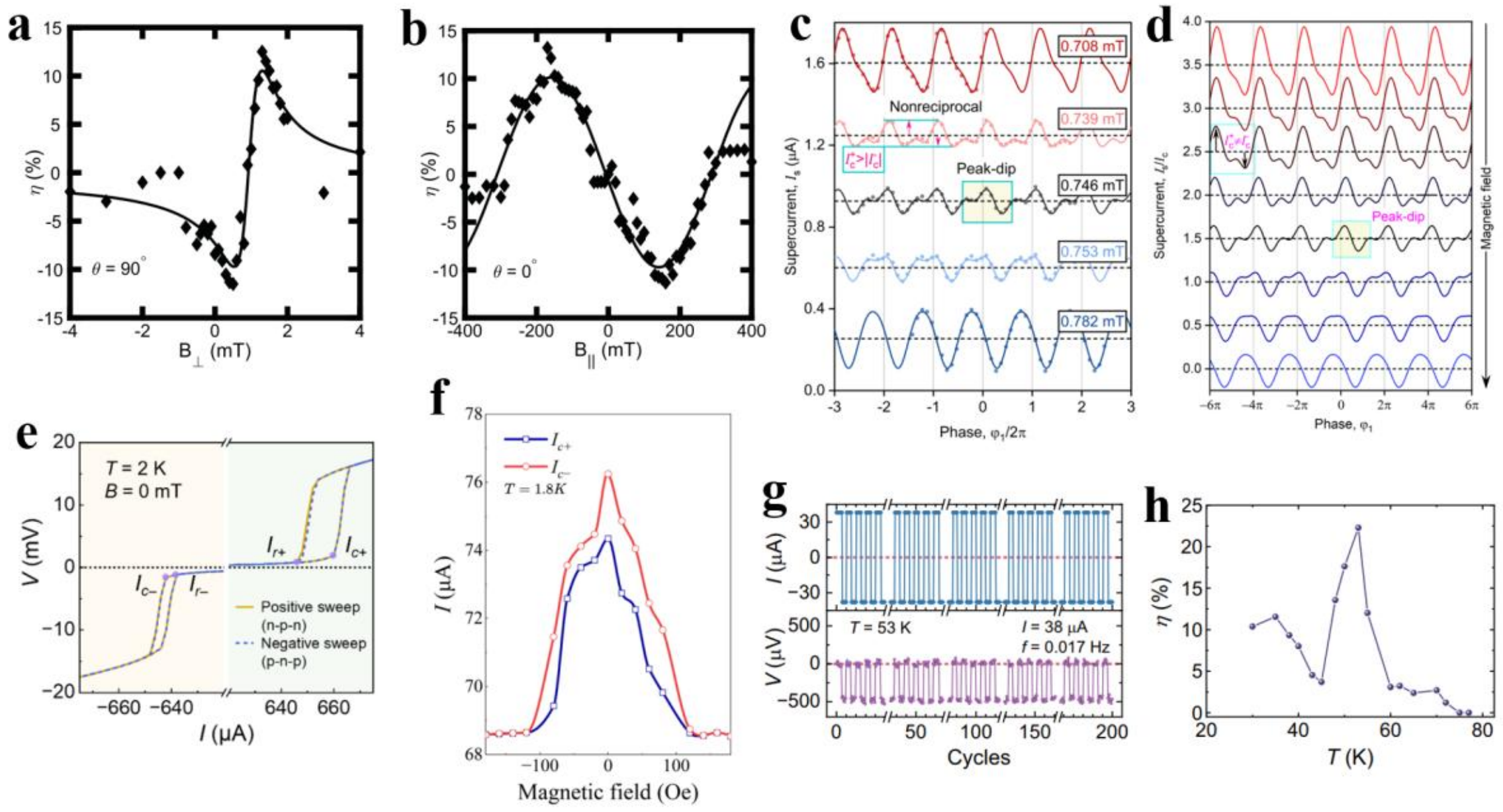

**Fig.3** a) Gaussian-like SDE efficiency $\eta$ induced by out-of-plane magnetic field in $NbS_2/NbSe_2$ heterojunction. b) Sinusoidal SDE efficiency $\eta$ induced by in-plane magnetic field in $NbS_2/NbSe_2$ heterojunction. c) Current-phase relation of a $NbSe_2/Bi_2Se_3$ SQUID at different magnetic fields. d) Simulation results of the CPR in Fig.3c without taking account of Majorana bound states. e) V-I characteristics in FeSe nanosheets upon forward and reversal sweepings. f) Critical currents versus the out-of-plane magnetic field in a 45°-twisted FeSe device. g) Rectification effects over long-term cycles in a BSCCO device. h) The diode efficiency versus temperature in the BSCCO device in Fig.3g. Panels (a-b) are reproduced with permission from [61], Copyright (2026), The Authors, some rights reserved; exclusive licensee Springer Nature, CC BY-NC-ND 4.0. Panels (c-d) are reproduced from [67], CC BY 4.0. Panel (e) is reproduced from [69], CC BY 4.0. Panel (f) is reproduced from [70], CC BY 4.0. Panels (g-h) are reproduced with permission from [71], Copyright (2025), The Authors, some rights reserved; exclusive licensee Springer Nature, CC BY-NC-ND 4.0.

## 4.2 Few-layer devices

Reducing materials to few-layer thickness often enhances tunability and exposes new symmetry-breaking channels. Wang et al. [72] achieved nonvolatile electrical control over the sign of nonreciprocal charge transport in few-layer ferroelectric polar metal $WTe_2$. By exploiting the intrinsic coexistence of switchable out-of-plane ferroelectricity and magnetochiral anisotropy, reversing the ferroelectric polarization flipped the slope of the second-harmonic magnetoresistance versus magnetic field, and this switched state remained after removing the external biasing field. The nonreciprocity was associated with a Drude-like mechanism enhanced by a large orbital magnetic moment arising from interband coherence near the Fermi level, countering the common view that Drude-type responses are trivial and devoid of band-geometric information—a perspective that may prompt re-evaluation of other "intrinsic" assignments [73].

Bauriedl et al. [74] reported an SDE in few-layer $NbSe_2$ under an out-of-plane field with $\eta$ exceeding 60%. An in-plane field perpendicular to the current did not drive the SDE but broke the antisymmetry of efficiency versus out-of-plane field. The out-of-plane field driven diode behavior was attributed to the valley-Zeeman spin–orbit interaction intrinsic to $NbSe_2$, which couples the out-of-plane field to the valley degree of freedom, simultaneously breaking TRS and IS [75]. This mechanism is distinct from the Rashba-type effects dominant in many other systems, illustrating the diversity of symmetry-breaking routes in layered dichalcogenides.

Varshney and Agarwal [76] theoretically demonstrated that longitudinal nonreciprocal charge transport can occur in noncentrosymmetric conductors even when preserving TRS, via disorder-induced asymmetric scattering (specifically skew scattering and side-jump).

They proposed an enhancement of $\eta$ in gated bilayer graphene near the Lifshitz transition, consistent with an experimental report of $\sim 30\%$ [77]. This theoretical work broadens the conceptual landscape by showing that TRS breaking, while sufficient, is not strictly necessary for nonreciprocal transport—a point that may have been overlooked in many experimental interpretations.

# 5. High-$T_c$ superconductors

For huge application prospects, high temperature superconductors have long been in the center of condensed matter study, mostly for rising up the $T_c$ and exploring a general mechanism of superconductivity, leaving both research and applicable potentials for nonreciprocal transport.

## 5.1 Iron-based superconductors

The iron-based superconductors have been a particularly contentious arena for SDE studies, largely due to the potential for thermoelectric artifacts to mimic genuine nonreciprocal signals. In bulk FeSe single crystals, Terashima et al. [13] observed nonreciprocal transport signals but excluded an intrinsic symmetry-breaking origin, instead attributing them to thermoelectric effects. By systematically correlating second-harmonic resistance with contact resistance and exchanging current/voltage electrode configurations, they revealed that the signals were artifacts sensitive to local heating and thermal gradients. This finding serves as an important warning: in measuring second-harmonic resistance for determination of inversion-symmetry breaking, one should rigorously rule out Joule heating influences, especially in semimetals and narrow-gap semiconductors. Concurrently, Nagata et al. [14] reported a zero-field SDE in FeSe flakes and attributed it to the same thermoelectric mechanism—Joule heating at contacts generating an in-plane temperature gradient due to shape asymmetry, with the large Seebeck coefficient of FeSe converting this gradient into an additional current. A separate heater was placed on the wider side of FeSe sample offering an artificial heating gradient. When the heater was off, the thinner side gathered more heating and locally higher temperature due to lower dissipation, driving a thermal contribution to the currents, making $I_{c+} > I_{c-}$. When heater was on, the heat gradient was reversed and so were the critical currents $I_{c-} > I_{c+}$. This interpretation was also supported by their control experiments where asymmetric-shaped NbN films with a smaller thermal effect showed no SDE at all.

In contrast to these thermoelectric interpretations, a following work by Han et al. [69] on FeSe nanosheets claimed an observation of intrinsic SDE in devices of both asymmetric and symmetric shapes. Notably, the V–I characteristics in Fig.3e of forward and reversal sweepings coincided perfectly with each other over long-term cycles, with no hysteresis or drift as signatures of thermal effect. A spontaneous JDE was also observed in Fig.3f in a 45°-twisted FeSe system after eliminating thermal influence [70]. This disagreement with some studies attributing signals to thermoelectric artifacts and others to intrinsic effects, suggests that multiple mechanisms may coexist in the same system, and that the experimental standards for distinguishing intrinsic from extrinsic signals need to be sharpened. It also deepens our understanding of what constitutes truly intrinsic nonreciprocity, and how to design control experiments that can definitively rule out thermal artifacts.

## 5.2 Cuprates

The high-transition-temperature cuprate superconductors offer the prospect of SDE at elevated temperatures, which is of significant practical interest. As shown in Fig.3g and 3h, Qi et al. [71] realized a stable zero-field SDE in $Bi_2Sr_2CaCu_2O_{8+\delta}$ (BSCCO) flake devices up to 72 K, with the highest diode efficiency $\eta \sim 22\%$. Tian et al. [78] subsequently provided macroscopic nonreciprocal transport evidence of both in-plane and out-of-plane inversion-symmetry breaking in BSCCO, arising from subtle lattice distortions. More impressively, in Josephson junctions made of twisted Bi2212 and Pb-Bi2212 flakes, Wang et al. [9] demonstrated a zero-field SDE with perfect $100\%$ efficiency. After a current training that temporarily implants a directional bias into the junction at 70K, a zero-field perfect SDE was observed where $I_{c+}$ vanished and $I_{c-}$ remained more than 0.3mA at a microwave irradiation power of 38dBm. Moreover, the voltage outputs were quantized Shapiro steps $V = n \cdot hf/2e$ and highly tolerant to fluctuations. The perfect SDE was realized in other devices of different twisted angles or tunneling area, in the highest temperature of 83K. This work represents a milestone in the field, achieving the ideal diode limit at liquid-nitrogen temperatures, and suggests that cuprates may offer a practical path toward high-temperature superconducting rectifiers.

## 5.3 Nicklates

Discovered more recently, nickelate superconductors present high transition temperatures under high pressure or in thin films, but nonreciprocal transport features have not been widely reported in this system until very recently. A study by Nishida et al. [79] made advances in $SrTiO_3$/(La,Sr)$NiO_2$ bicolor and $SrTiO_3$/$LaAlO_3$/(La,Sr)$NiO_2$ tricolor superlattices. A clear B-linear $R_{2\omega}$ signal was observed in the superconducting state for both superlattices, whereas in the normal state only the tricolor stack showed a weak slope. Scanning transmission electron microscopy directly revealed asymmetric residual apical oxygen distributions in bicolor stacks (one layer at top and two layers at bottom interfaces) and in tricolor stacks (up to three layers at the top interface). While this local interfacial oxygen nonuniformity primarily contributed to the enhancement of the superconducting state, the global stacking symmetry was also proposed as the key factor for the normal-state signal unique to the tricolor cell. This work highlights that in newly discovered superconductors, even subtle structural asymmetries at the atomic level can give rise to nonreciprocal transport, and that systematic stacking engineering may be a powerful tool for enhancing such effects.

# 6. Topological and magnetic materials

As counterparts of superconductors, topological and magnetic materials provide an alternative way to study the direction-selective currents that do not rely on superconductivity but general nontrivial band topology and magnetism within.

## 6.1 TI thin films

Topological insulators, with their spin-momentum-locked surface states, provide a natural arena for nonreciprocal transport, as the surface Dirac fermions inherently couple current and spin degrees of freedom. Ahn et al. [80] reported nonreciprocal charge transport in $Bi_2Te_3$ films up to room temperature, induced by the Rashba effective field, with a maximum $\gamma \sim 3.89 \times 10^{-3} A^{-1}T^{-1}$ at 100 K and a high Rashba parameter of $16.3\ eV \cdot Å$ extracted from anisotropic magnetoresistance curves. In a nanowire geometry, Nikodem et al. [81] observed SDE in $BiSbTeSe_2$ nanowires with efficiency $\eta \sim 0.3$, as displayed in Fig.4a and simulated in Fig.4b, where the critical current showed periodic oscillations with parallel field at the flux quantum $\Phi_0 = h/2e$. This finding provided direct evidence of interference between top and bottom surface channels in a nano-SQUID model incorporating skewed current-phase relations. In dual-gated Sn-$Bi_{1.5}Sb_{0.8}Te_2S$ devices, Li et al. [82] found quantum Hall (QH)-mediated nonreciprocal transport with $\gamma \sim 2.26 \times 10^5 A^{-1}$ near QH plateau transitions, attributed to asymmetric scattering between chiral QH edge states and disorder-broadened Landau orbits [83]. The fact that $R_{2\omega}$ vanishes in well-developed plateaus indicates that the intrinsic QH bulk contribution is not involved, distinguishing this effect from quantum anomalous Hall [84] or bilinear magnetoelectric mechanisms. These topological-insulator studies collectively demonstrate that the surface states can yield substantial nonreciprocal signals, but the underlying mechanisms—Rashba fields, quantum interference, or edge-state scattering—vary considerably with device geometry and doping.

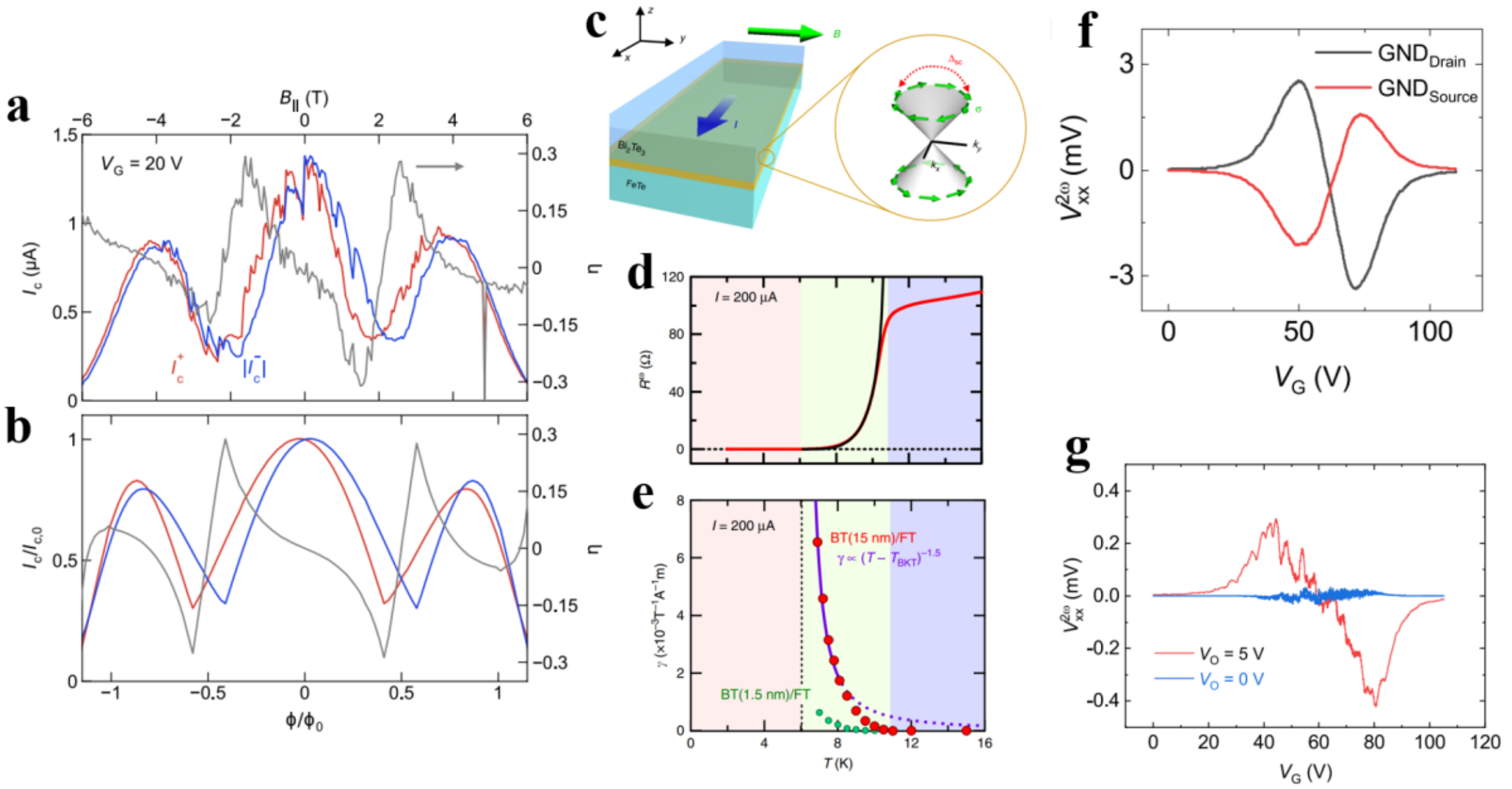

**Fig.4** The experimental a) critical currents and diode efficiency of $BiSbTeSe_2$ nanowire and their simulations b) from a tight-binding model. c) The structure of topological insulator $Bi_2Te_3$/FeTe heterojunction. There is a coexistence of spin-momentum-locked topological surface state and superconducting state at the interface. d) The RT curve (red) of $Bi_2Te_3$/FeTe and a BKT fitting (black). e) The nonreciprocal efficiency $\gamma$ versus temperature, with the red/green dots representing 15nm/1.5nm-thickness $Bi_2Te_3$ films in

Fig.4c. f) Second-harmonic longitudinal responses as a function of gate voltage $V_G$ for grounding drain and source in a 6-SL $MnBi_2Te_4$ device. g) Second-harmonic longitudinal responses versus gate voltage $V_G$ with a small AC current, measured with and without a superimposed AC gate voltage $V_0$ via a transformer in the same 6-SL $MnBi_2Te_4$ device. Panels (a-b) are reproduced from [81], CC BY 4.0. Panels (c-e) are reproduced from [85], CC BY 4.0. Panels (f-g) are reproduced with permission from [86], Copyright (2024), The Authors, some rights reserved; exclusive licensee Springer Nature, CC BY-NC-ND 4.0.

## 6.2 TI/SC heterojunctions

Integrating topological insulators with superconductors introduces additional symmetry-breaking channels and proximity effects. In $Nb/Bi_{0.8}Sb_{1.2}Te_3/Nb$ Josephson junctions, Behner et al. [87] detected a Josephson diode effect with maximum $\eta \sim 7\%$, which is modest compared to theoretical predictions of $20-40\%$ [88, 89]. The authors inferred ballistic transport channels and skewed current-phase relations in the presence of half-integer Shapiro steps as the mechanism. This quantitative discrepancy between experiment and theory suggests that disorder, interface quality, or finite-temperature effects may significantly suppress the ideal diode performance, and systematic tuning of junction length, carrier density, or interface quality is desirable to clarify the role of ballistic channels versus disorder-induced effects.

Fig.4c shows a heterojunction structure of $Bi_2Te_3$ and FeTe, where superconductivity can be induced at the interface. The superconductivity is confirmed by the RT curve in Fig.4d. Finding the response could be nonreciprocal, Yasuda et al. [85] reported a dramatic increase in $\gamma$ from $6\times10^{-8}A^{-1}T^{-1}m$ in the normal state [90] to $6.5\times10^{-3}A^{-1}T^{-1}m$ in the superconducting transition. As illustrated in Fig.4e, the $\gamma$ value diverges as temperature approaches the BKT transition and can be suppressed by reducing $Bi_2Te_3$ thickness, while at higher fields, $R_{2\omega}$ reverses sign strictly for in-plane field alignment. When FeTe was replaced by the non-magnetic superconductor $PdTe_2$, Masuko et al. [91] reported a similar emergence of reciprocal signals near the superconducting transition, with $\gamma \sim 6.8\,A^{-1}T^{-1}m$. Differently, a sign reversal was observed under tilted fields and interpreted as a vortex-type crossover from spontaneous to field-induced behavior. The similarities and differences between these two heterostructures—both involving $Bi_2Te_3$ but with different superconducting partners—highlight the sensitivity of nonreciprocal transport to the detailed nature of the interface and the superconducting pairing symmetry [92].

## 6.3 Magnetic TIs

Although not intrinsically superconducting, the antiferromagnetic topological insulator $MnBi_2Te_4$ has drawn considerable attention as a platform for exploring the interplay between topology, magnetism, and nonreciprocal transport—a combination that is highly relevant for future superconducting hybrid devices. Wang et al. [93] reported both a nonlinear anomalous Hall effect and a diode-like longitudinal response in even-layer $MnBi_2Te_4$ ($\gamma \sim 7 \times 10^{-11} m^2 A^{-1}$), attributed to the quantum-metric dipole under broken symmetry. Both longitudinal and transverse second-harmonic conductivities appear to be independent of the square of longitudinal conductivity, that is, independent of the scattering time by simple Drude model. This scattering-time-independent nonlinear conductivity, along with observed sign reversal upon switching the AFM order, disappearance above $T_N$, explicitly ruling out Berry-curvature dipole and skew-scattering contributions through scaling analysis. In contrast, Xu et al. [86] demonstrated in Fig.4f and 4g that much of the second-harmonic signal in similar devices can be artificially induced by gate-voltage oscillations arising from large AC currents, with sign and magnitude changing drastically upon switching the grounded electrode—an effect reproduced by intentionally superimposing an AC gate voltage. This raises a serious caution: even well-designed experiments may contain artifacts that mimic intrinsic nonreciprocal signals. An earlier work by Zhang et al. [94] on 5-SL and 4-SL $MnBi_2Te_4$ showed a nonreciprocal signal with $\gamma \sim 4.5 \times 10^3 A^{-1}$, which was edge-selective, magnetically switchable, gate-voltage controllable, and vanished in even-layer (4-SL) devices at zero magnetization, attributed to asymmetric dispersion of chiral edge states hybridized with trivial edge channels. Taken together, while several scenarios have been proposed to account for nonreciprocal transport in $MnBi_2Te_4$, the possible contamination from gate-voltage-induced artifacts suggests that a definitive assignment of the intrinsic mechanism may require further investigation—and perhaps a standardized measurement protocol to separate genuine signals from experimental artifacts.

Ye et al. [95] reported nonreciprocal transport in $MnBi_2Te_4$/Pt bilayers only below $T_N$, with a $\pi/2$ phase offset between longitudinal and transverse responses and non-monotonic field dependence confirming magnon-mediated asymmetric scattering of spin-momentum-locked surface electrons [96]. Meanwhile, the disappearance of signals with field in xz- or yz-planes and current along x-direction ruled out spin–orbit torques or Rashba effects as the origin. The contrasting results across different $MnBi_2Te_4$-based devices in which some emphasizing quantum-metric dipoles, magnon scattering or raising artifact concerns, underscore the need for cross-validation using multiple measurement geometries and complementary techniques.

## 6.4 Non-superconducting magnetic systems

Although not superconductors, magnetically ordered systems with broken inversion and time-reversal symmetry provide important comparative platforms for understanding nonreciprocal transport mechanisms. Kar et al. [97] discovered both nonlinear Hall effect and nonreciprocal transport in ferromagnetic Weyl metal $SrRuO_3$ thin films. The transverse signal followed a sinusoidal dependence on lattice angle and scaled linearly with bias current, consistent with the Berry curvature dipole. Distinctly, the longitudinal signal exhibited an orthogonal angular dependence to the transverse signal, suggesting one-dimensional chiral edge modes [84]. Notably, the nonlinear Hall field ratio was $1.7 \times 10^{-7} m/V$, four orders larger than the bulk quantum metric dipole effect [98], indicating that surface or edge contributions can dominate in thin-film geometries.

In semimetal $T_d$-$MoTe_2$, Hu et al. [99] observed a nonlinear planar Hall effect and separated it into intrinsic Berry-connection-polarizability and extrinsic nonlinear Drude-like terms [100, 101]. The two tensor elements for orthogonal current directions carried opposite signs, and for each, the intrinsic and extrinsic contributions also opposed each other. Counterintuitively, the extrinsic term grew faster at low temperatures as scattering weakened, so the quantum-geometric response remained hidden as a small offset and could only be extracted by scaling analysis. This implies that cleaner samples may actually obscure, rather than reveal, the intrinsic contribution, which is a counterintuitive but important insight for the design of experiments aimed at detecting quantum-geometric effects.

Masuda et al. [102] reported nonreciprocal transport in the centrosymmetric helimagnet $YMn_6Sn_6$ below room temperature, with a sharp and critically enhanced peak at the helimagnetic-to-antiferromagnetic phase boundary, as illustrated in Fig.5a and Fig.5b. This was explained as asymmetric electron scattering off ferromagnetic fluctuations or clusters, combined with spin–momentum locking in the helical state. Concurrently, a related study [103] on the same compound focused on the transverse conical phase under a perpendicular magnetic field. Unlike the first work, the resistivity in the ordered transverse conical state was primarily due to band-asymmetry effects rather than scattering, as evidenced by the finite nonreciprocal signals within the transverse conical (TC) state, although similar peaks were also found near the proper screw (PS) boundaries (Fig.5c). Through angle-resolved measurements of the nonreciprocal resistivity in Fig.5d, two coexisting contributions were identified as conventional magnetochiral anisotropy and a toroidal moment. These contrasting observations indicate that the microscopic origin of nonreciprocal transport in this system is configuration-dependent, varying with field orientation and magnetic phase, and a unified understanding awaits further investigation.

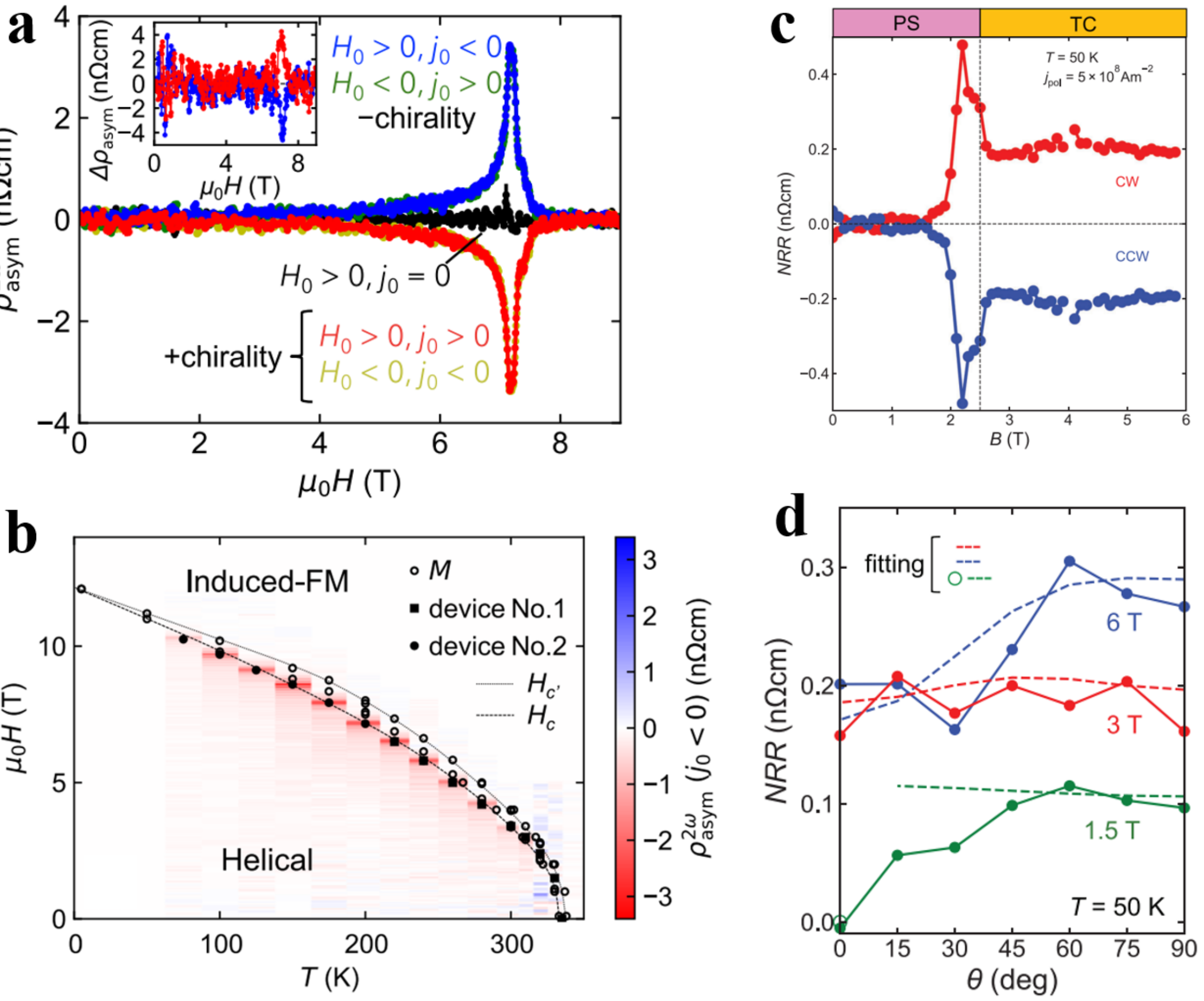

**Fig.5** a) The second-harmonic resistivity versus the magnetic field parallel to the current along c-axis in a $YMn_6Sn_6$ device at 200K. b) The field-temperature diagram for the second-harmonic resistivity in devices of Fig.5a. c) Nonreciprocal response versus the magnetic field perpendicular to the current along c-axis in a $YMn_6Sn_6$ device from a separate report. d) Nonreciprocal response versus the angle of magnetic field in device of Fig.5c. At 0 or 90 degrees the field is along the a- or c- axis. Panels (a-b) are reproduced from [102], CC BY 4.0. Panels (c-d) are reproduced from [103], Copyright (2025), The Authors, published by Wiley-VCH GmbH, CC BY-NC-ND 4.0.

A recent work [104] reported nonreciprocal transport in the zigzag antiferromagnet $NdRu_2Al_{10}$ under zero external magnetic field below $T_N$. The crystal contains a zigzag chain of Nd atoms, where the two sublattices experience opposite local electric-field gradients along the b-axis due to IS broken. In a simple 4-electrode device, nonreciprocity emerged below $T_N$ and the coefficient $\gamma$ increased at lower temperature, indicating intrinsic correlations to antiferromagnetic orders. In another device with multiple voltage channels, the sample was found to have domains with different AF orders. At both ends of the device, domain A had a (↑↓↑) polarity and domain B had an opposite polarity of (↓↑↓), and the two polarities coexisted in the middle. To notice, $\gamma$ showed opposite signs but equal intensity at domains A and B and it varied smoothly with spatial position. However, its magnetic-field response remained even-symmetric—sharply contrasting with conventional field-induced rectification effects. The effect was attributed to a magnetic toroidal dipole order that simultaneously broke TRS and IS, with the anomalously large magnitude ascribed to a strong internal effective field generated by c–f exchange coupling between conduction electrons and localized Nd moments [105]. This work demonstrates that antiferromagnets with toroidal order can host zero-field nonreciprocal transport, offering an alternative to ferromagnetic or superconducting platforms [106].

Multiferroic (Ge,Mn)Te thin films [107] exhibited a pronounced nonreciprocal resistance with $\gamma' = 1.3 \times 10^{-11}\, m^2 A^{-1} T^{-1}$, five orders larger than in nonmagnetic GeTe [108]. Yoshimi et al. attributed this to inelastic magnon-mediated scattering in a spin–momentum locked Fermi surface, supported by a characteristic hump in temperature-dependent $R_{2\omega}$ and its field-sensitive peak shift. The remarkably large magnitude in this system suggests that magnon scattering can be a highly efficient mechanism for generating nonreciprocal signals, potentially rivaling or exceeding those from purely electronic band-geometric effects.

## 7. Conclusions and perspectives

In conclusion, the research on nonreciprocal charge transport has grown quickly in diverse quantum materials, yet a clear and unified picture remains elusive. What emerges from this survey is not a single dominant mechanism, but a rich interplay of band-structure effects, vortex dynamics, scattering processes, and thermoelectric artifacts that often coexist within the same material family. The superconducting diode effect seems a delicate but informative probe: it can arise from intrinsic symmetry breaking, but equally from asymmetric pinning, geometric design, or simple Joule heating.

Several themes stand out. First, the distinction between intrinsic and extrinsic contributions remains a central challenge. Control experiments such as swapping voltage contacts, varying device geometry, or checking thermal gradients are becoming essential for credible claims. Second, the field is moving beyond the traditional requirement of simultaneous inversion and time-reversal symmetry breaking, with recent work showing that disorder, magnetic domains, or even trapped flux can generate diode-like responses under milder conditions. Third, high-temperature operation in cuprates and the demonstration of perfect efficiency mark important milestones, though practical applications will require better stability, higher critical currents, and scalable fabrication.

Looking ahead, we see three promising directions. One is systematic comparative studies across closely related compounds, which can isolate the role of specific electronic or structural parameters. Another is the integration of nonreciprocal elements into functional devices, where gate tunability and nonvolatile switching are particularly attractive. A third is the continued refinement of theoretical frameworks that can handle coexisting mechanisms on equal footing, rather than treating them in isolation. Perhaps most importantly, the community would benefit from establishing standardized protocols for distinguishing genuine nonreciprocal signals from artifacts—an approach that promises both speed and reliability in future developments.

## Acknowledgements

This work was supported by the National Natural Science Foundation of China (Grants No. 12274369 and No. 12604087) and by Zhejiang Provincial Natural Science Foundation of China (Grants No. LZ25A040003 and No. LQN26A040014). Y. H. acknowledges the support from the China Postdoctoral Science Foundation (Grant No. 2025M773352).